# Origin of Martian Moons from Binary Asteroid Dissociation

Geoffrey A. Landis
NASA John Glenn Research Center
Mailstop 302-1
21000 Brookpark Road, Cleveland, OH 44135
geoffrey.a.landis@nasa.gov

**Abstract**

The origin of the Martian moons Deimos and Phobos is controversial. One hypothesis for their origin is that they are captured asteroids, but the mechanism requires an extremely dense martian atmosphere, and the mechanism by which an asteroid in solar orbit could shed sufficient orbital energy to be captured into Mars orbit has not been well elucidated.

Since the discovery by the space probe Galileo that the asteroid Ida has a moon "Dactyl", a significant number of asteroids have been discovered to have smaller asteroids in orbit about them. The existence of asteroid moons provides a mechanism for the capture of the Martian moons (and the small moons of the outer planets). When a binary asteroid makes a close approach to a planet, tidal forces can strip the moon from the asteroid. Depending on the phasing, the asteroid can then be captured.

Clearly, the same process can be used to explain the origin of any of the small moons in the solar system

## What is the origin of the moons of Mars?

The origin of the Martian moons Deimos and Phobos is controversial.

Joseph Burns, in "Contradictory Clues as to the Origin of the Martian Moons" [1992], states:

*"The scientific jury remains divided on the question of… satellite origin. Dynamicists argue that the present orbits could not have been produced following capture."*

Two theories of formation are:
1. Deimos and Phobos formed along with Mars
2. Deimos and Phobos are captured asteroids

While Deimos and Phobos appear asteroid-like, capture requires loss of energy of the objects as they drop into Mars's gravity well. The obvious candidate is atmospheric braking, but this needs an implausibly thick atmosphere around Mars, and requires fine-tuning the atmosphere to be dense enough for the excess energy to be bled off in atmospheric friction, but not so dense that the resultant orbits decay quickly. Further, low densities (Deimos $\rho=1.7$, Phobos $\rho=1.7$) indicate that Deimos and Phobos are not solid rocky objects, and may not have the structural strength to have survived an aerocapture.

Thus, while it appears that Deimos and Phobos were objects captured from the asteroid belt, the aerodynamic friction argument for how they were captured does not seem reasonable. The key question is thus:

*How were Deimos and Phobos captured?*

## Asteroid moons

Recently it has been recognized that many asteroids are binary objects. The first asteroid confirmed to have a moon [Belton *et. al* 1996] was the asteroid 243 Ida, observed in 1993 by the spacecraft Galileo to have a moon (tagged "Dactyl"). Since 1999, radar and optical observations have detected a large number of asteroids with moons. It is now accepted that moons of asteroids are not uncommon. Asteroid moons are as large as 80 km diameter for the components of the double asteroid 90 Antiope [Merline *et al*. 2000]. It is notable that the sizes of discovered asteroid moons brackets the size of the moons of Mars.

Since it is relatively easy to dissociate an asteroid from its moon, over the lifetime of the solar system many such binary pairs must have been dissociated, and in the early solar system the number of asteroids with moons must have been greater than it is now.

## Asteroid moons: table

| Asteroid | Estimated size of moon | Moon discovered |
|---|---|---|
| 243 Ida | 1.6 km | 1993 |
| 3671 Dionysus | <1km | 1997 |
| 45 Eugenia | 15 km | 1999 |
| 1996 FG3 | | 1999 |
| 90 Antiope | 80 km | 2000 |
| 762 Pulcova | 15 km | 2000 |
| 2000 DP 107 | 0.5? km | 2000 |
| 107 Camilla | 10 km | 2001 |
| 1999 KW-4 | 0.4 km | 2001 |
| 2000 UG11 | | 2001 |
| 87 Sylvia | 13 km | 2001 |

## Gravitational Capture

In this paper, I suggest that the moons of Mars are captured asteroids.

As evidenced by the heavy bombardment era of the early solar system, near the beginning of its existence the solar system was occupied by a much larger number of asteroids and planetisimals than are now seen. Most likely some of these objects were in orbits with periods very close to Mars, and that early Mars would occasionally experience fly-by of objects at close, but not impacting, range. As the solar system evolved into its present state, objects with initial orbits close to Mars would have been gravitationally scattered away from this location. In the early solar system, some asteroids must have had initial orbits that with a hyperbolic excess relative to Mars of very close to zero.

Recognizing that many of these objects must have been binary objects, a mechanism is proposed for the capture of the moons of Mars: by tidal dissociation of an asteroid moon from its parent body.

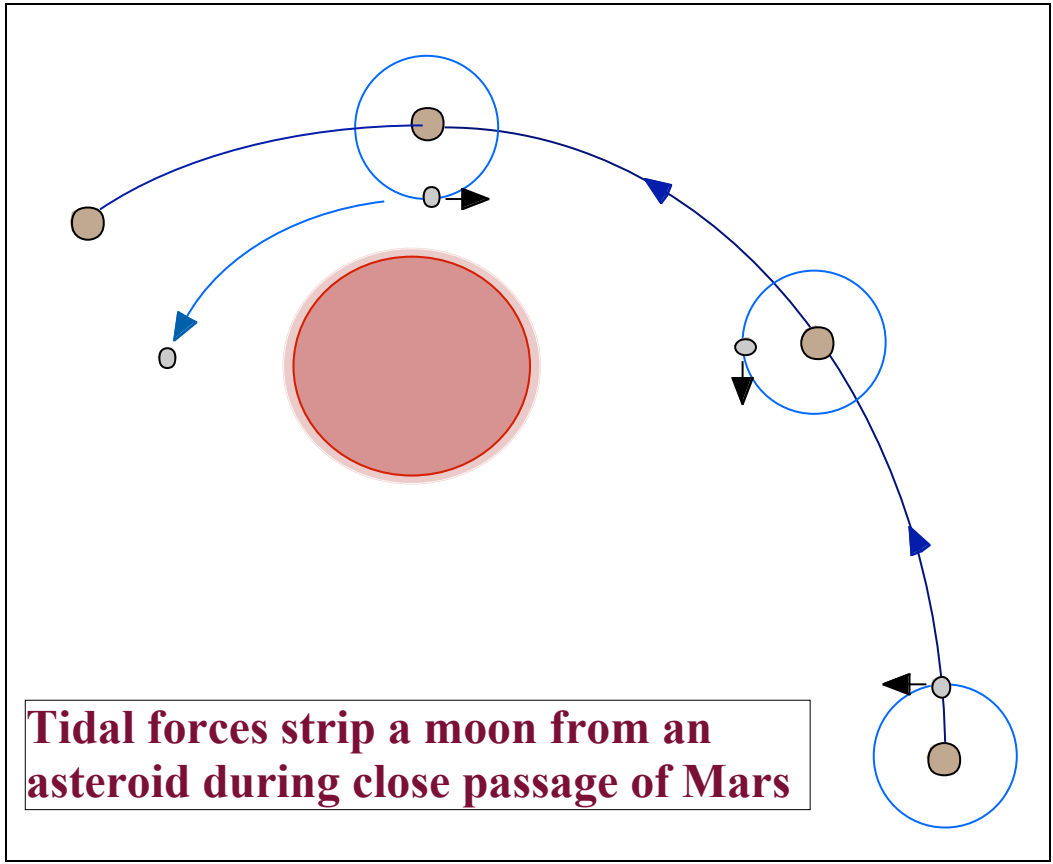

**Tidal forces strip a moon from an asteroid during close passage of Mars**

Stripping of a moon from an asteroid is similar to the classic problem of Roche disruption of a cohesionless body, with the additional constraint that the moon has a kinetic energy that is equal to half of the gravitational binding energy. Tidal forces lower the gravitational potential energy barrier, and dissociation of the moon from its primary will occur when the barrier is lowered so that the barrier can be overcome by kinetic energy of the moon.

Binding force: $$F_{binding} = G \frac{m_{asteroid} m_{moon}}{r_{orbit}^2}$$

Tidal force: $$F_{tidal} = \frac{dF_{mars}}{dr} \Delta r = -2G \frac{M_{mars} m_{moon}}{r_{approach}^3} \Delta r$$

Here $r_{orbit}$ is the unperturbed orbital distance of the asteroid moon from the parent asteroid, $r_{approach}$ is the distance from Mars at which gravitational dissociation occurs, and $\Delta r$ is the distance of the asteroid moon from the asteroid, expressed as a variable.

The maximum of the potential barrier occurs at the point where the tidal force equals the binding force:

$$\Delta r = \left(\frac{m_{asteroid}}{2M_{mars}}\right)^{1/3} r_{approach}$$

Binding potential energy: $E_{binding} = -G\dfrac{m_{asteroid}m_{moon}}{r_{orbit}}$

(effective) tidal potential energy: $E_{tidal} = -G\dfrac{M_{mars}m_{moon}}{r_{approach}^3}\Delta r^2$

The barrier is lowered by an amount:

$$\Delta E = -0.63 G \frac{m_{moon}}{r_{approach}}\left(M_{mars} m_{asteroid}^2\right)^{1/3}$$

Since the orbital energy is half of the escape energy, escape occurs when ΔE equals the orbital energy. This will occur at an approach distance

$$r_{approach} = 1.26 \left(\frac{M_{mars}}{m_{asteroid}}\right)^{1/3} r_{orbit}$$

As the asteroid comes closer to Mars than this dissociation distance, the potential barrier lowers rapidly with decreasing distance. When the asteroid reaches the Roche limit, the potential barrier disappears.

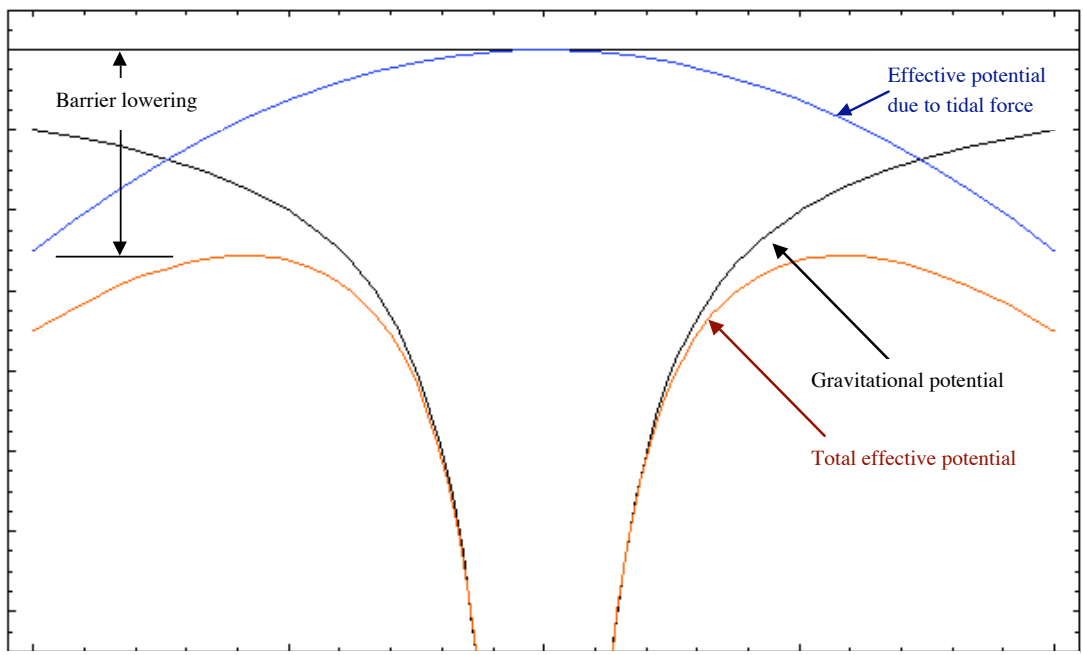

Figure: potential barrier of asteroid and tide

## Example case:

For the worst (closest to Mars) case, assume that Phobos originated as a moon of a Ceres-sized asteroid:

$M_{asteroid} = 8.7 \; 10^{20}$ kg $= M_{mars}/738$

Assume a low orbit, $r_{orbit} = 500$ km

Then $r_{approach} = 1.26/738^{1/3}$ (500 km) = 5700 km

This is two thirds of a Mars radius above the surface of Mars.

Assuming that the dissociation occurs at the periapsis of a parabolic ($V_\infty=0$) orbit, the velocity at periapsis is

$$v_{periapsis} = \sqrt{2\frac{GM_{mars}}{r_{approach}}} = 3.88 km/\sec$$

If the parent asteroid is smaller than Ceres, or if the initial moon orbit is higher, the dissociation occurs further from Mars, and the periapsis velocity is smaller; if the hyperbolic excess velocity $V_\infty$ is higher, the periapsis velocity will be greater.

## Resulting Orbit

When a binary asteroid dissociates, conservation of energy requires that the resulting orbits of the two fragments must sum to a value lower in energy by an amount equal to the original gravitational binding energy. The energy can be removed from the trajectory of either the asteroid, the moon, or both. If the energy removed from the moon's resulting trajectory is greater than the hyperbolic excess energy, the moon will be captured.

The energy of moon's resulting trajectory after dissociation will depend on whether the moon is closer or further from Mars than the parent body when it dissociates, the position of the moon in its orbit, and the orientation of the orbital plane of the moon/asteroid system compared to the orbital plane of the motion about Mars; it will also depend on whether the barrier is lowered over a time constant that is small or large compared to the orbital period of the moon. For the best case, at dissociation the orbital velocity of the moon will be directed opposite to the orbital motion of the asteroid/moon binary. In this case, as the asteroid approaches the Roche limit the moon can have an effective increment in velocity equal to its orbital velocity.

For the example case of a Ceres-sized primary asteroid, the orbital velocity of a low orbit is approximately 0.347 km/s. Subtracting this from the periapsis velocity of 3.88 km/s at $r_{approach}$ = 5700 km, the velocity at periapsis after dissociation will be 3.53 km/s. The resulting orbit will be captured, but will be highly elliptical.

## Orbital Eccentricity

The preceding calculation suggests that binary asteroid dissociation can cause capture of an asteroid, but the resulting orbit is highly eccentric, quite unlike the nearly-circular orbits of present-day Deimos and Phobos. However, circularization of eccentric orbits can occur by the process of viscoelastic tidal damping, a process well discussed in the literature. One characteristic of viscoelastic damping is that the damping process can decrease the eccentricity and the energy of the orbit, but the angular momentum of the orbit is unchanged by the process. (This neglects several higher-order effects. Over long periods of time the orbital energy can be changed by tidal drag, or by interactions between several moons, but for the present calculation, these effects will be ignored).

It is instructive to compare the angular momentum of the captured moon in the example case to the angular momentum of Phobos. 3.53 km/s periapsis velocity times periapsis distance of 5700 km gives an angular momentum (per unit mass) of 20,100 km$^2$/s for the captured-asteroid case, nearly identical to the 19,250 km$^2$/s angular momentum of Phobos. After circularization, the orbit of the captured asteroid moon would be very nearly identical to (slightly higher than) that of Phobos. It thus seems plausible that the moons of Mars may be captured by the means proposed.

## Roche Capture

These example calculations have been for the case of an orbiting binary asteroid.

An alternate possibility is that the binary asteroid disrupted is a contact binary, where two parts are structurally distinct but in contact, rather than orbiting. Asteroid Castalia may be an example of a contact binary. In this case, the dissociation at periapsis is by the classic Roche-limit effect, and conservation of energy requires that the resulting orbit has an orbital energy that is lower than the initial orbit by an amount equal to the gravitational binding energy of the pair. This will result in capture if the initial hyperbolic excess energy is less than the gravitational binding energy. Since even the largest asteroids have a relatively low binding energy, only a very loose capture is possible, and it is difficult to account for Deimos or Phobos in this way.

## Conclusions

It has been shown that dissociation of a binary asteroids can result in capture of one of the objects as a moon, but is this a likely enough process to account for the moons of Mars? It is clear from the cratering record that early in its history many asteroidal-sized objects impacted on Mars. A much larger number must have approached close to, but missing, Mars. Since it is now known that many asteroids have moons, and presumably many of the asteroids that made close approaches to Mars were binary asteroids, it is reasonable to phrase the question in the opposite sense: how likely could it be that *none* of the near-misses of asteroids flying past Mars resulted in capture of a moon?

The example calculation assumed a hyperbolic excess velocity $V_\infty$ of zero: the initial orbit was very close (in DV) to the orbit of Mars. Capture is unlikely if the hyperbolic excess velocity is very much larger than zero; a moon orbiting a main-belt asteroid, for example, would not be likely to be captured regardless of how close an encounter is made. Since the inner edge of the asteroid belt is most likely defined by asteroids ejected from the belt by gravitational encounters with Mars, however, it is likely that in the early solar system, asteroids with near-Mars orbits may have been common.

The present analysis has centered on the origin of the moons of Mars, however, a similar origin for other small moons in the solar system can be proposed.

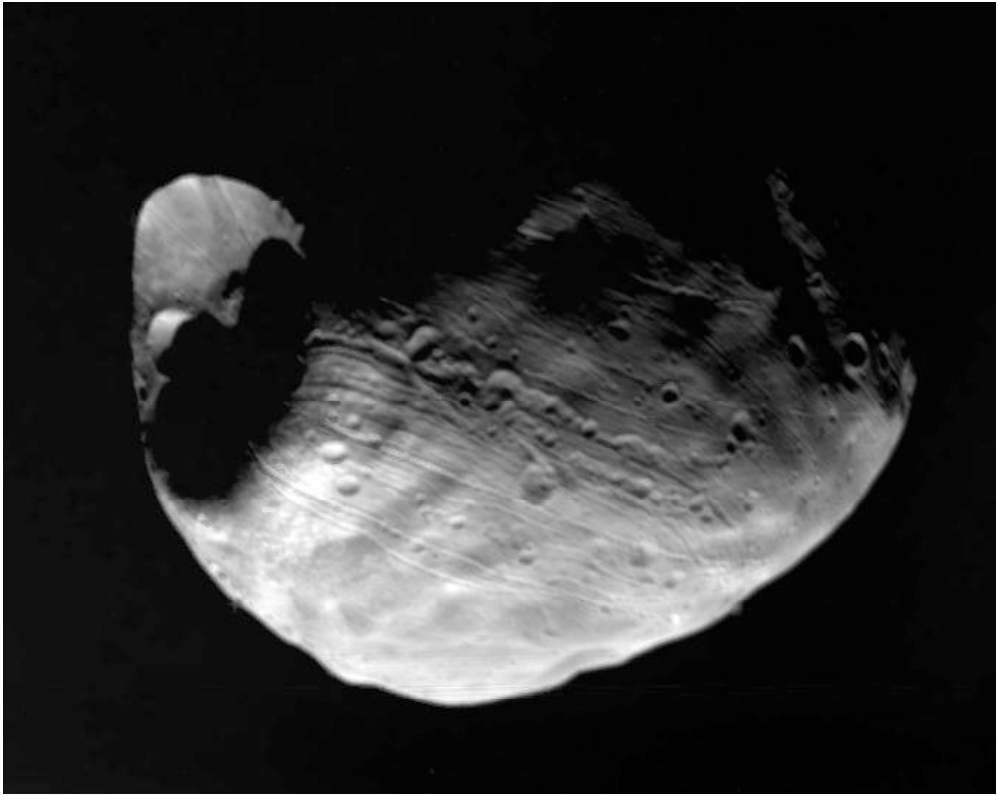
Phobos: image from Viking spacecraft, courtesy NASA/JPL

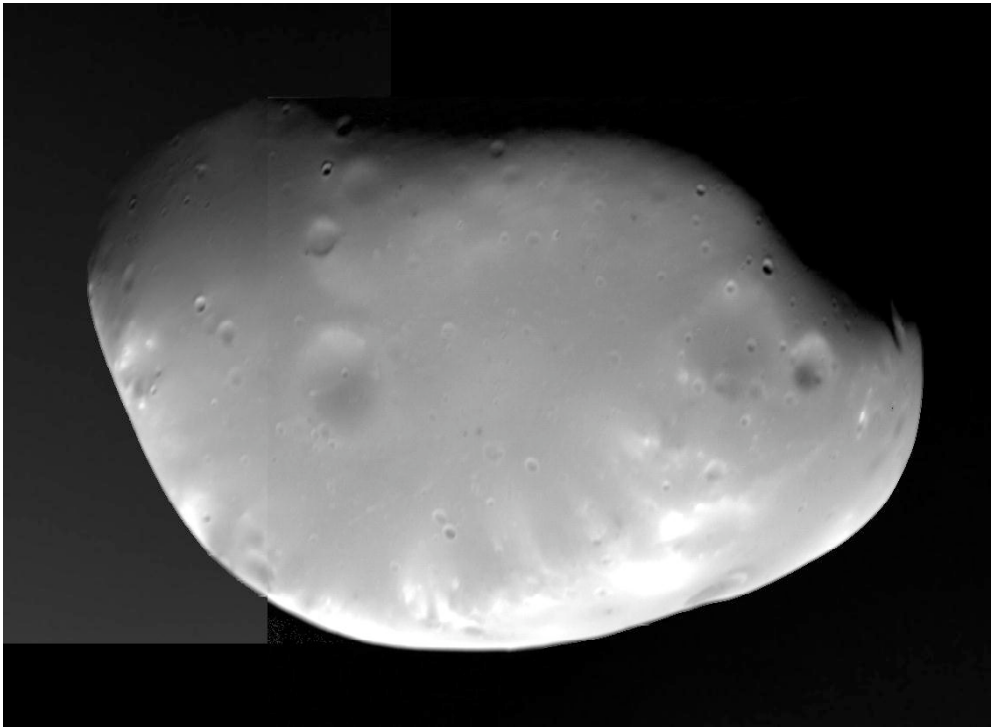
Deimos: mosaic image from Viking spacecraft, courtesy NASA/JPL

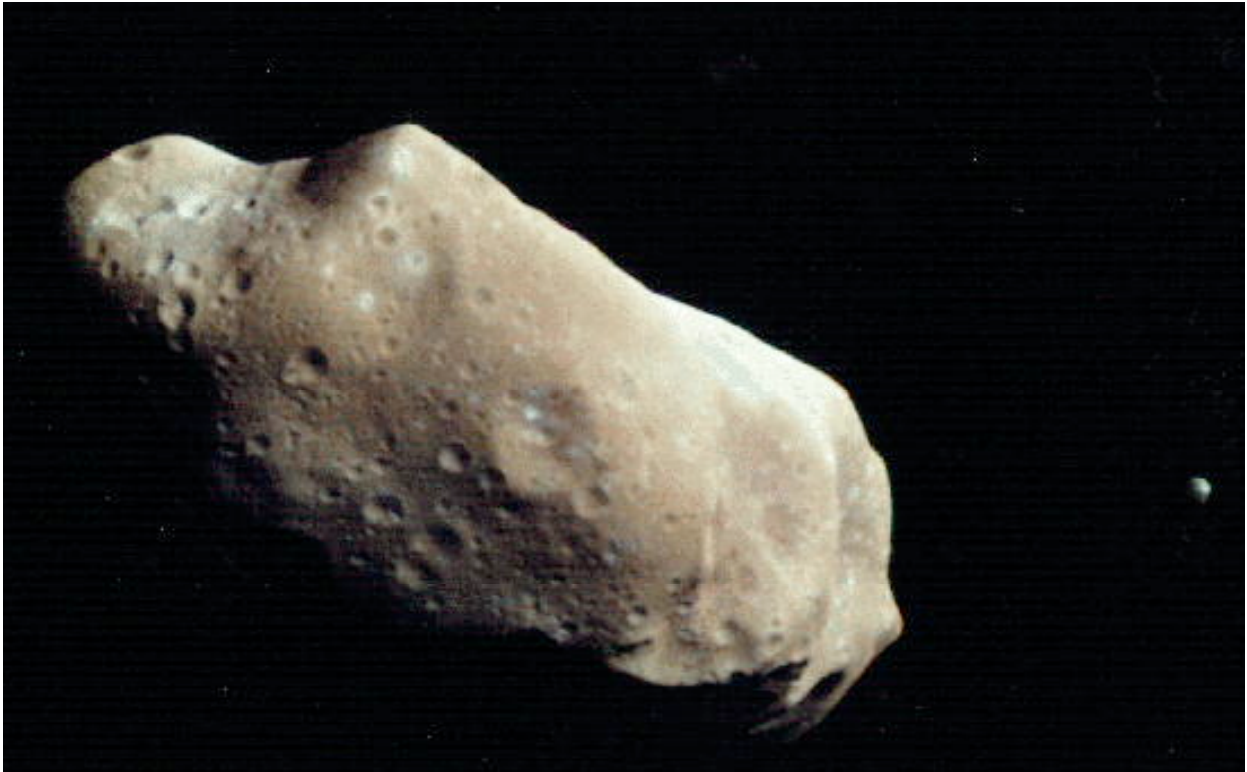
Asteroid Ida and moon Dactyl: image from Galileo spacecraft, courtesy NASA/JPL